# ZEUS-2: a second generation submillimeter grating spectrometer for exploring distant galaxies

Carl Ferkinhoff*[a], Thomas Nikola [a], Stephen C. Parshley [a], Gordon J. Stacey [a], Kent D. Irwin [b], Hsiao-Mei Cho [b], Mark Halpern [c]

[a]Department of Astronomy, Cornell University, Ithaca, NY, USA 14853
[b]NIST Boulder, Boulder, CO USA 80305-3337
[c]Department of Physics and Astronomy, University of British Columbia, Vancouver, B.C., V6T 1Z1, Canada

## ABSTRACT

ZEUS-2, the second generation (z)Redshift and Early Universe Spectrometer, like its predecessor is a moderate resolution (R~1000) long-slit, echelle grating spectrometer optimized for the detection of faint, broad lines from distant galaxies. It is designed for studying star-formation across cosmic time. ZEUS-2 employs three TES bolometer arrays (555 pixels total) to deliver simultaneous, multi-beam spectra in up to 4 submillimeter windows. The NIST Boulder-built arrays operate at ~100mK and are readout via SQUID multiplexers and the Multi-Channel Electronics from the University of British Columbia. The instrument is cooled via a pulse-tube cooler and two-stage ADR. Various filter configurations give ZEUS-2 access to 7 different telluric windows from 200 to 850 micron enabling the simultaneous mapping of lines from extended sources or the simultaneous detection of the 158 micron [CII] line and the [NII] 122 or 205 micron lines from z = 1-2 galaxies. ZEUS-2 is designed for use on the CSO, APEX and possibly JCMT.

## 1. INTRODUCTION

ZEUS-2, is designed to study the star formation history of the Universe from early times until the present epoch continuing the work begun with its predecessor, ZEUS.[1] ZEUS-2 is optimized for the detection of faint, broad submillimeter and redshifted far-IR (FIR) lines such as the ground state fine-structure lines from $C^0$, $C^+$, $O^0$, and the mid-J rotational lines from CO. These lines are important coolants, in some case the dominate coolant, of major phases of the ISM. They are among the brightest lines from a galaxy. By comparing their relative strengths knowledge of both the excitation source and the physical characteristics (i.e. temperature, density) of the gas can be deduced.

Observations with ZEUS on CSO have produced the first extragalactic detection of the $^{13}CO(6-5)$ line[2], the first detection of the [OIII] 88 micron line[3] at redshift greater than 0.05, and the first detection of the 158 micron line[4] of [CII] from between z = 1 – 2. In total ZEUS has made 12 detections of the [CII] 158 micron line in a z = 1 – 2 galaxy survey[5]. A mid-J CO survey of local infrared bright (IRB) and ultra luminous IR galaxies (ULIRG) has also been performed to probe the nature of the activity in nearby galaxies[6]. ZEUS-2 will continue and build upon this important and ground breaking work of understanding star formation across cosmic time by working towards three scientific goals:

1. ***Investigate star formation in the early Universe*** by detecting redshifted far-IR fine-structure lines from distant (z > 0.25 – 5) galaxies. ZEUS-2 is capable of detecting the 158 micron [CII] line from hundreds of SCUBA and Spitzer sources, thousands of sources likely to be detected in Herschel's survey, and any high redshift ULIRG detected by the WISE survey. The [CII] line traces the physical extent of a starburst as well as its intensity thereby making the study of the [CII] line in the early universe critical for understanding the star formation history of the Universe.

2. ***Measure redshifts of optically obscured distant sources*** using the [CII] 158 micron line. While there is a wealth of sources detectable with ZEUS-2 that already have spectroscopic redshifts, several current missions (e.g. Herschel and WISE) will produce source catalogs containing many infrared bright galaxies with only photometric redshifts. These sources, in addition to many sources from SCUBA and IRAS catalogs lacking spectroscopic redshifts, are prime targets for ZEUS-2 as they are likely enshrouded by dust making optical spectroscopic follow-up difficult if not impossible. ZEUS-2 is uniquely suited for this task due to its large instantaneous bandwidth (~5% of a band) and large redshift range accessible across its seven bands.

*cferkinh@astro.cornell.edu; phone 1 (607) 255-5891; fax 1 (607) 255-3433; submm.astro.cornell.edu

3. ***Investigate star formation and molecular gas excitation in nearby galaxies*** by simultaneously mapping CO(7-6), $^{13}$CO(6-5), [CI] 370 and 609 micron, and [NII] lines to produce maps of the physical conditions of major components of the ISM. These maps will allow us to study how galaxy dynamics (i.e. density waves, resonances, etc.) interact with the ISM and stimulate star formation. Additional high-J CO and $^{13}$CO observations will distinguish between cosmic rays, turbulence, and/or X-rays as the dominate source of heating of molecular clouds in ULIRGS.

To achieve these goals, ZEUS-2 takes the optical design of ZEUS and marries it to several technological advances that push its capabilities significantly further. Namely, ZEUS-2 requires no liquid cryogens, significantly reducing its operating costs and allowing the instrument to be used at sites with better atmospheric properties. ZEUS-2 also features three state-of-the-art TES sensed bolometer arrays with tuned back-shorts allowing for ~90% quantum efficiencies across 7 telluric windows from 200 to 850 microns. The arrays also allow for observations between 5 and 10 spatial positions on the sky and the simultaneous detection of five important submillimeter lines from extended sources or two lines from z = 1 - 2 and 2.70 - 3 galaxies. These key improvements make ZEUS-2 more sensitive than its predecessor by factors of 1.35 and 1.6 at 350 and 450 microns respectively (ZEUS-2's primary bands), increase the accessible redshifts by more than a factor of 2, and produce a 10x improvement in mapping speed. Compared to other submillimeter spectrometers, ZEUS-2 is the most sensitive submillimeter spectrometer. In its primary bands ZEUS-2 is 3 to 5 times more sensitive than the best heterodyne receivers and 10 times more sensitive than the Herschel spectrometers.

## 2. SCIENCE

### 2.1 High-Redshift Universe

Multi-wavelength studies have shown that the star formation rate (SFR) was higher in the early universe and peaked between a redshift of 1-3, or between 2 – 6 billion years after the Big Bang[7]. The peak SFR per unit co-moving volume then was 10 to 30 times higher than at present. The majority of the star formation likely occurred at z > 1 in ultra-luminous infrared galaxies (ULIRG, $L_{FIR} > 10^{12} L_\odot$), which due to their high dust content, emit most of their radiation at far-infrared wavelengths[8]. The local universe also contains ULIRGs but it is not clear how the two populations (high and low z) are related and if they have similar properties.

#### 2.1.1 High-z Spectral Probes

The primary tools for probing the early universe with ZEUS-2 are the far-IR fine-structure emission lines that have been redshifted into the submillimeter. These lines arise in different phases of the ISM (e.g. ionized gas: [OIII], [NII]; atomic gas: [CII], [OI]) based on the ionization potentials of their parent species. The far-UV (FUV) photons required for ionization arise from young stars or AGN. These same photons are responsible for the photo-ejection of electrons from the surface of dust grains, which in turn heat the gas and collisionally excite the far-IR lines. Many of the lines are extinction free and usually optically thin, allowing the emitted photons to escape, and the gas to cool so it can collapse and form stars. The lines are therefore excellent probes for the density and the temperature of the gas from which they were emitted as well as the strength of the energizing and/or ionizing UV fields.

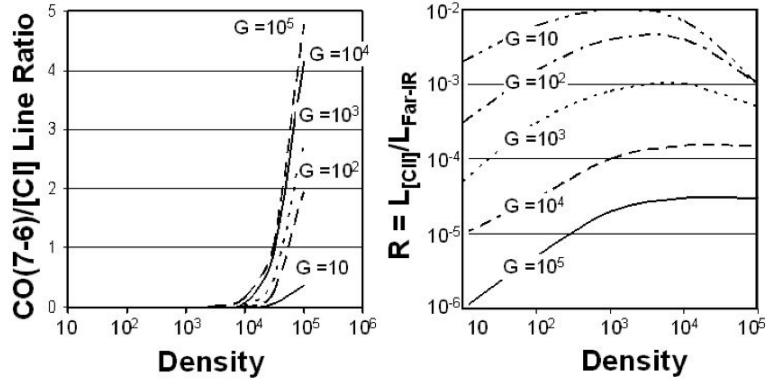

Figure 1: (left) CO(7-6)/[CI] line ration as a function of density and FUV field and (right) R, [CII] luminosity to FIR luminosity ratio as a function of density and FUV field (from Kauffman et al.[9]).

In some cases, even the detection of a single line can constrain internal properties of a galaxy. This is the case of the 158 micron line of [CII] that dominates the cooling in photo-dissociation regions (PDRs) and atomic clouds. It is also an important coolant of the warm ionized medium. Typically the brightest FIR line from a galaxy, with line luminosity between 0.001 and 1% of the galaxies FIR continuum, it arises in PDRs on the surface of moderately dense molecular clouds of $n\sim10^{2.5}$ and $10^{-4}$ cm$^{-3}$.[10] For such cloud densities, the ratio R=[CII]/FIR has a strong inverse dependence on the strength of the interstellar FUV-radiation field, G (Fig. 1). Measurements of this ratio can constrain both the source of the UV photons as well as the physical extent of the emitting region.[11] Large R ($10^{-3}$ - $10^{-2}$) indicates modest G as is found for normal galaxies like the Milky Way. Moderate R ($10^{-3}$) indicates stronger FUV fields as found in starburst galaxies like M82, while a small R ($10^{-4}$) indicates very strong fields associated with intense and compact star forming regions like Orion[12], the starbursts of local ULIRGs or regions close to an AGN. R can also indicate the age of a starburst as a younger starburst will contain more massive stars produce a more intense FUV field, and hence a lower G. As the starburst ages, the most massive star will die causing the strength of the FUV to decrease and resulting in a higher R.

The [CII] 158 micron line is the primary spectral probe for ZEUS-2 due to the strength of the line and its utility in understanding the physical conditions of the gas and FUV fields. However, to gain a complete picture of star formation in the early Universe, additional probes are needed, and for this ZEUS-2 can turn to a large set of other FIR fine-structure lines (see Table 1). For instance, the ratio of the [OI] 63 micron line to [CII] is sensitive to gas pressure and enables the determination of the gas density in the PDR. Fine structure lines from $N^+$ and $O^{++}$, which are found only in HII regions, are important probes of the gas density and radiation fields in those regions. Using [NII] and [OIII] lines it is possible to determine the amount of [CII] emission arising from the ionized medium as well as the hardness of the radiation.[11,13,14] Additionally, these lines are often the next bright lines from a star-forming galaxy after [CII].[15,16,17]

Table 1: ZEUS-2 Extragalactic Probes

| Species | Transition | E.P.[a] | λ(μm) | A(s$^{-1}$) | N$_{crit}$(cm$^{-3}$) |
|---|---|---|---|---|---|
| $O^0$ | $^3P_1$ - $^3P_2$ | 228 | 63.184 | $9.0\times10^{-5}$ | $4.7\times10^5$ |
| $O^{++}$ | $^3P_2$ - $^3P_1$ | 440 | 51.815 | $9.8\times10^{-5}$ | $3.6\times10^3$ |
|  | $^3P_1$ - $^3P_0$ | 163 | 88.356 | $2.6\times10^{-5}$ | 510 |
| $C^+$ | $^2P_{3/2}$ - $^2P_{1/2}$ | 91 | 157.741 | $2.1\times10^{-6}$ | $2.8\times10^3$ |
| $N^+$ | $^3P_2$ - $^3P_1$ | 188 | 121.898 | $7.5\times10^{-6}$ | 310 |
|  | $^3P_1$ - $^3P_0$ | 70 | 205.178 | $2.1\times10^{-6}$ | 48 |
| $C^0$ | $^3P_2$ - $^3P_1$ | 63 | 370.415 | $2.7\times10^{-7}$ | $1.2\times10^3$ |
|  | $^3P_1$ - $^3P_0$ | 24 | 609.135 | $7.9\times10^{-8}$ | $4.7\times10^2$ |
| $^{12}$CO | J=13-12 | 503 | 200.272 | $2.2\times10^{-4}$ | $2.5\times10^6$ |
|  | J = 9 - 8 | 249 | 289.120 | $7.3\times10^{-5}$ | $8.4\times10^5$ |
|  | J = 8 – 7 | 199 | 325.225 | $5.1\times10^{-5}$ | $5.9\times10^5$ |
|  | J = 7 - 6 | 155 | 371.651 | $3.4\times10^{-5}$ | $3.9\times10^5$ |
|  | J = 6 - 5 | 116 | 433.556 | $2.1\times10^{-5}$ | $2.6\times10^5$ |
| $^{13}$CO | J = 8 - 7 | 111 | 340.181 | $4.5\times10^{-5}$ | $1.7\times10^5$ |
|  | J = 6 - 5 | 55 | 453.498 | $1.9\times10^{-5}$ | $2.3\times10^5$ |

[a]Excitation potential: energy (K) of upper level above ground. [b]Molecules: Collision partner $H_2$ (100 K). Atoms: [CI], [OI] (H atoms), [NII], [OIII] (electrons).

### 2.1.2 High-z Science

Using ZEUS, Hailey-Dunsheath et al.[4] detected the [CII] 158 micron line from MIPS J142824.0+352619 and determined that vigorous star formation was occurring over several kilo-parsecs. The [CII] to FIR ratio for MIPS J142824 is $\sim 10^{-3}$, indicating that high-redshift ULIRGs appear as scaled up versions—in both star formation rate and physical size of the starburst—of local star forming galaxies with moderate far-UV fields (like M82). It appears that high-z ULIRGs are not like their low redshift counter parts which have compact star forming nuclei of a few hundred parsec, $L_{[CII]}/L_{[FIR]} \sim 10^{-4}$ and intense far-UV fields. Recent Herschel detections of the [OI] 63 micron and [OIII] 52 micron lines from MIPS J1428 and a [CII] detection from z = 2.3 submillimeter galaxy SMM J2135-0102 support this conclusion.[18,19] While it is clear that high redshift ULIRGs dominated by star formation appear as scaled up versions of local star-forming galaxies, the picture becomes less clear when AGN dominated and composite systems are considers.

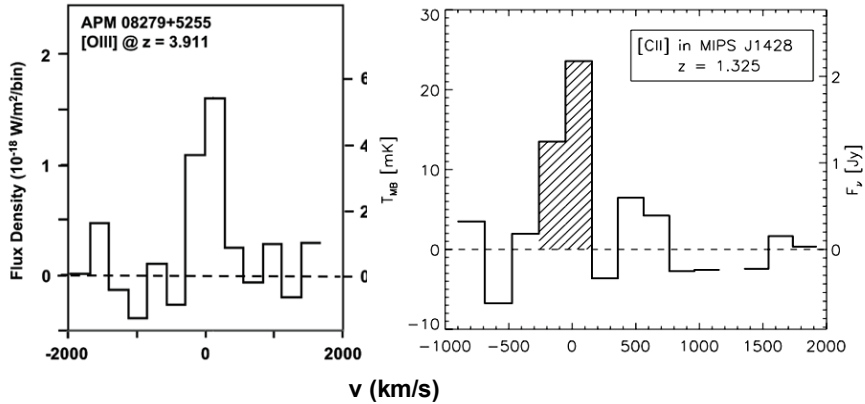

Figure 2: (left) [OIII] 88 μm line from APM 08279 and (right) [CII] 158 μm from MIPS J1428.[3,4]

Recently submitted work by Stacey et al.[5] describes the results of the ZEUS [CII] line survey in which the [CII] line is detected from 12 galaxies of a 13 galaxy sample. The heterogeneous sample consists of equal parts star formation dominated, AGN dominated, and mixed or poorly characterized systems. This survey shows a clear dichotomy between the star formation and AGN dominated system; star forming systems are similar to MIPS J1428 and local star forming galaxies with R $\sim 10^{-3}$ whereas the AGN dominated systems have R $\sim 10^{-4}$, similar to local ULIRGs. However, both systems show similar kilo-parsec scale emitting regions. Stacey et al. suggest a simple evolutionary model to account for similar sized emission region but difference in R. A merger first triggers an AGN phase that in turn triggers a galaxy-wide starburst with very intense FUV field produced from bright O/B stars such that R$\sim 10^{-4}$. Gradually the starburst ages and the intensity of the FUV field subsides along with the activity of the AGN resulting in the moderate intensity FUV field and R$\sim 10^{-3}$ of the star-formation dominated sample.

In the context of this model the dichotomy between high-z AGN and star-forming galaxies is simply a result of observing the same sample of galaxies, but at different stages in their evolution. Stacey et al. caution that the model is very speculative, however the recent ZEUS made detections of the [OIII] 88 micron line from two high-z AGN/starburst composite systems, combined with the high-z Herschel detections of the [OIII] 52 micron line suggest that an AGN does in fact have a younger more intense starburst than the purely starburst dominated systems in the early Universe. ZEUS-2 will provide more detections of the [CII] line and other FIR lines from fainter sources than ZEUS and over a wider range of redshift to test the current AGN/starburst/R correlations and Stacey et al.'s evolutionary model.

### 2.2 Local Universe

The discovery of ULIRGs by the IRAS mission was one of the mission's most exciting results. Among the most powerful sources in the local universe, their energy usually results from major merger of two similar sized galaxies. This merger, by compressing the ISM, triggers a starburst or causes accretion onto a central black hole forming an AGN that results in extreme FIR luminosities. Studies with ISO and Spitzer suggest that ¾ of these galaxies are powered by starburst, the remaining are powered by AGN. With ~30% of the molecular ISM in starburst galaxies being warm and dense[20,21,22,23], it is important to understand the interplay between star formation and this molecular gas. There are also implications for our understanding of star formation at early epochs in the Universe. Specifically, as mentioned above, local ULIRGs have [CII] to FIR ratios that are significantly smaller than local normal and star-forming galaxies. This can be caused by a [CII] deficit due to higher FUV fields, or because there is an additional FIR source not related to star formation and PDRs. Understanding the source of this deficit is clearly needed if [CII] is to be used as a probe of star formation.

### 2.2.1 Low-z Spectral Probes

The [CI] and CO lines are the dominate coolants in the interiors of molecular cloud. Measuring their strength probes directly the physical state of the gas. Specifically, the mid-J CO lines accessible to ZEUS-2 trace warm, dense molecular gas associated with dense PDRs, X-ray dissociation regions (XDRs) or shocks. Observing the run of line intensity as a function of J constrains the physical properties of the gas including temperature, density, CO fraction, and mass. The fine structure lines of neutral carbon, being well mixed in cloud interiors, are good traces of the cloud mass.[9,24] By observing both the 370 micron and 609

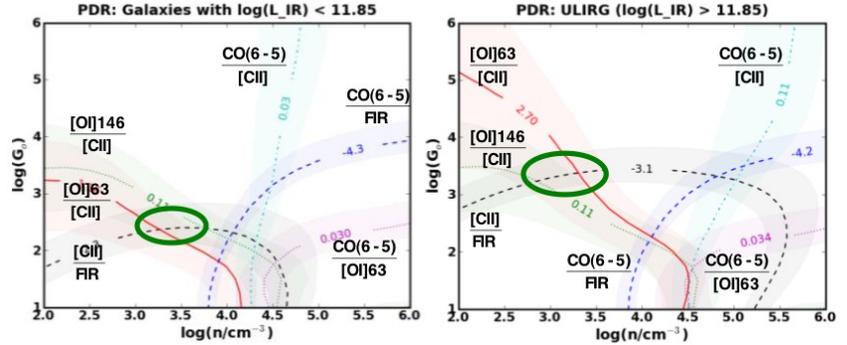

Figure 3: PDR plot of gas density versus FUV ($G_0$) (K1), together with contours of intensity ratios [CII], [OI] 63 and 146 μm, FIR continuum (from literature), and ZEUS/CSO CO (6-5). For the 17 IRB galaxies (left), and 6 ULIRGs (right) a single component (green circle) fits all data except the CO(6-7) line. A second component can produce a PDR fit for all lines from the IRB galaxies, but not for the ULIRGs. The ULIRGs therefore require an additional source of molecular cloud heating[6].

micron [CI] line the temperature of the gas can be determined since both lines are easily thermalized and usually optically thin. Combing both [CI] and CO detections is particularly valuable as they are respectively temperature and density sensitive. For example the ratio of the [CI] 370 micron line to CO(7-6) line is a sensitive probe of density and can reveal dense star forming regions (Fig. 1)

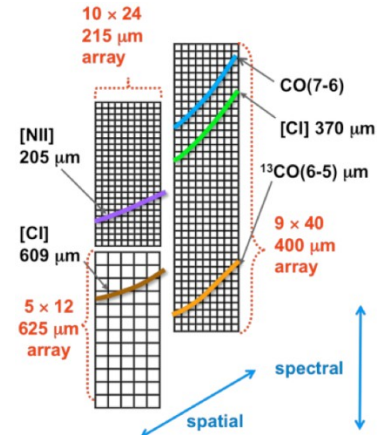

Figure 4: Schematic of ZEUS-2 arrays showing simultaneous detection of 5 important spectral probes.

### 2.2.2 Low-z Science

ZEUS has been very successful in probing the properties of nearby galaxies. As described in Hailey-Dunsheath et al. 2008[2], the first extragalactic detection of the $^{13}CO(6-5)$ line was detected from NGC 253 using ZEUS. From this detection, combined with ZEUS made detections of other mid-J CO lines and the [CI] 370 micron line, it was determined that gas in NGC 253 is warm (T~110K) and dense (~$10^4$ cm$^{-3}$) requiring the gas to be heated from turbulent decay in the molecular cloud or by cosmic rays. Both of these heat sources are likely to be driven by the formation of stars. Continuing this work and seeking to understand the source of the [CII] deficit in ULIRGs, a ZEUS survey of 27 nearby galaxies in the [CI] 370 micron, CO(6-5), CO(7-6) and CO(8-7) lines has been performed. Nikola et al.[6] (in prep) describes the result of this survey of 20 infrared bright galaxies (IRB) and 7 ULIRGs. As shown in figure 3, by comparing the CO(6-5), [CII], [OI] and FIR emission from these galaxies Nikola et al. discovered, that the CO(6-5)/FIR and CO(6-5)/[OI] ratios are the same between to the two classes (IRB and ULIRG). For both classes a single PDR model cannot reproduce the observe intensities. A two component model fits the IRB galaxies reasonably well, but under-predicts the CO(6-5) emission from the ULIRGs by a factor of 6. The ULIRGs must have an additional source of heating not related to the PDR. Possible sources include XDRs around AGN, cosmic rays, and/or turbulence.[22,2] This may also explain the [CII] deficit in ULIRG systems.

ZEUS-2 will continue this study of nearby systems with the ability to simultaneously detect the CO(7-6), $^{13}CO(6-5)$, [CI] 370 and 609 micron, and [NII] lines (see Fig. 4). By mapping galaxies in these lines the physical conditions of the gas will be traced as it is compressed in spiral arms and stimulated to form stars. ZEUS-2 can also detect higher-J (e.g. J=9-8) that constrains the source excitation in the gas (i.e. PDR, turbulence, cosmic rays) to complete our understanding of the [CII] deficit in local ULIRGs.

### 2.3 Redshift Search

In order to constrain star-formation as a function of redshift it is important to obtain accurate redshifts of the many sources being identified in current ongoing and past infrared/submillimeter surveys. Searching for redshifts using the [CII] 158 micron line is compelling because of its relative strength compared to other lines (200 times brighter than the brightest CO lines, and typically the brightest of all FIR lines) and because high redshift source are likely to be very IR luminous so that the [CII] line will also be very luminous.

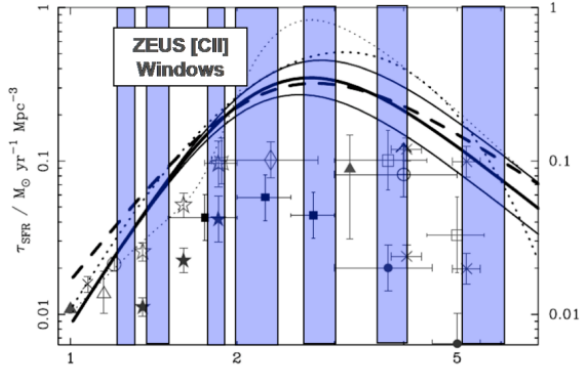

Figure 5: The co-moving star formation history of the Universe (S2). The original Madau plot based on optical/UV HDF observations are the filled marks.[25] Open symbols correct this data for dust extinction.[26] The 2 curves with error envelopes are models based on the SCUBA data. The transparent bars mark redshift ranges accessible in the 158 μm [CII] line to ZEUS-2.

ZEUS-2 is uniquely suited for performing a [CII] redshift search. With seven bands from 200 to 850 microns ZEUS-2 can detect the [CII] line from sources with redshifts between 0.25 and 5 with 50% coverage of that range (see Fig. 5). A source's redshift need only be known to ~5% because of ZEUS-2's large instantaneous bandwidth. The source list for a ZEUS-2 redshift search is subsequently quite large and can include sources from the Spitzer/SWIRE survey, SCUBA detected SMGs, IRAS sources, and sources from the Herschel HerMES survey.

Many of the sources from the catalogs listed above are found in well studied field. Arguably then the most exciting sources will be those identified by the WISE mission that is performing an all-sky IR survey at 3.3, 4.7, 12 and 23 microns.[27] WISE is ~1000 times more sensitive than IRAS and is expected to detect millions of sources, many of which will be ULIRGs and HyLIRGs ($L_{FIR} > 10^{13}$ $L_\odot$). WISE is able to detect any galaxy like F15307+3252— a z~1 HyLIRG with $\log(L_{FIR})$ = 13.12 $\log(L_\odot)$—out to a redshift of z = 3.[27] Assuming the [CII] to FIR ratio for these sources is $10^{-3}$, as has been observed by ZEUS, then ZEUS-2 can detect the [CII] line from any WISE ULIRG as long as the source falls within one of its 7 bands. These sources will provide an exciting new set of high redshift galaxies outside of the traditionally well studied fields (e.g. Lockman hole, HDF, CDF, etc.)

## 3. INSTRUMENT DESIGN

### 3.1 Optical Systems

ZEUS-2 will use the same echelle grating that is currently in ZEUS. This grating is an R2 echelle blazed at 359 micron in 5$^{th}$ order (blaze angle of 63.43° and groove spacing of 992 micron) and operated in Littrow mode. The telluric windows at 200, 230, 300, 350, 450, 625, and 850 microns are accessible in the 9$^{th}$, 8$^{th}$, 6$^{th}$ 5$^{th}$, 4$^{th}$, 3$^{rd}$, and 2$^{nd}$ orders of the grating respectively (see Fig. 12). The detector arrays give ~5% instantaneous coverage of a telluric window. Tilting the grating between 57 ° to 73 ° via a stepper motor allows access to rest of the window.

The rest of the ZEUS-2 optical system is very similar to ZEUS, but has been redesigned to maximize the usable field of view (FoV) within the limited instrument envelopes dictated by the CSO and APEX telescopes. The final design increases the usable FoV from 83" to 155", eliminates one mirror, and allows the grating to be tilted over a larger range. Figure 6 illustrates the design. A high-density polyethylene window accepts the f/12 telescope beam, which reaches a focus just inside the 3K stage. The flat mirror M1 then directs the beam to collimating mirror M1, through a Lyot stop, and to M3 changing the f/# from 12 to 2.75, appropriate for the detector pixel size. Next, the beam passes through the spectrometer entrance slit and expands to fill the right half of the collimating mirror M4. This creates a collimated 10 cm beam that illuminates the 38 cm long grating. The dispersed beam from the grating is directed back to the left side of M4 and then to the flat mirror M5 that sends the beam to the focal plane array.

Order selection is performed by band-pass filters mounted directly in front of the detector arrays. Several different filter configurations can be used to allow simultaneous detection of different FIR and submillimeter lines. The baseline configuration splits the 400 micron array (see Fig. 4) into separate 350 and 450 micron bands (i.e. 5$^{th}$ and 4$^{th}$ order), but allows for the simultaneous mapping of the 5 most important cooling/diagnostic lines. Other filter configurations enable simultaneously detecting the [CII] and the [NII] 122 or 205 micron lines over much of the redshift range z = 1 – 2, or [CII] and [OIII] 52 micron from z = 2.70 – 3.00. In addition to the order selecting filters, other filters—two submillimeter long pass (LP) filters, an optical/near-IR scatter filter, and an antireflection coated quartz filter—are used to intercept unwanted background radiation and prevent out of band leaks. This is important as the background optical power is very low (~1pw). The LP, BP, and scatter filters are sourced from P. Ade's group at Cardiff University. Where possible, ZEUS-2 uses the filters from ZEUS.

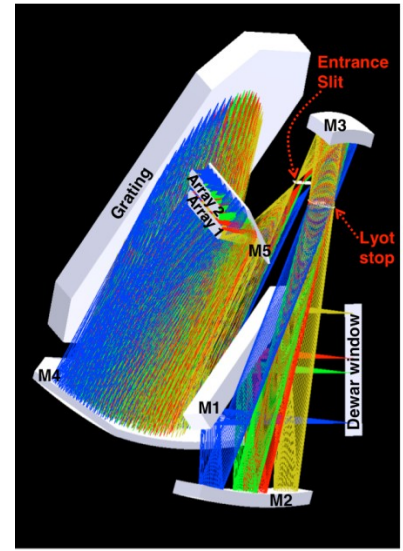

Figure 6: ZEUS-2 optics. Green and blue beams are spatial limits of Array 1, while yellow and red beams are spatial limits for array 2/3.

### 3.2 Dewar and Cryogenic Systems

The ZEUS-2 dewar is an all-aluminum vacuum shell designed in-house and manufactured locally. Figure 7 shows a picture of ZEUS as well as a schematic drawing of the ZEUS-2 internals. Great care was taken in minimizing the mass of the system in order to meet the tight weight limits inside the APEX receiver cabin. Cooling is provided by a Cryomech model PT407 pulse-tube cooler that provides 28 W and 300 mW of cooling power at 55 and 3 K respectively. A dual stage ADR from Janis provides 13 μW and 400 nW of cooling power at 1 K and 100 mK. An outer heat shield is held at 50 K and inner shield, along with the optics, are at 3K. This stage also serves as thermal dump for ADR cycles and thermal intercept for the ADR cold heads. A 1 K shield around the detector package is thermal attached the 1 K ADR stage, which also serves as a thermal intercept for the array wiring. The detector itself is held at 100 mK—its operating temperature—and enclosed in a 100 mK housing. A GRT temperature sensor on the detector package is used for monitoring the detector temperature, which is then fed into a PID loop that actively controls the current in the ADR magnet and keeps the detector temperature constant. The cooling power provided by the ADR allows the 100 mK operating temperature to be maintained for about one day.

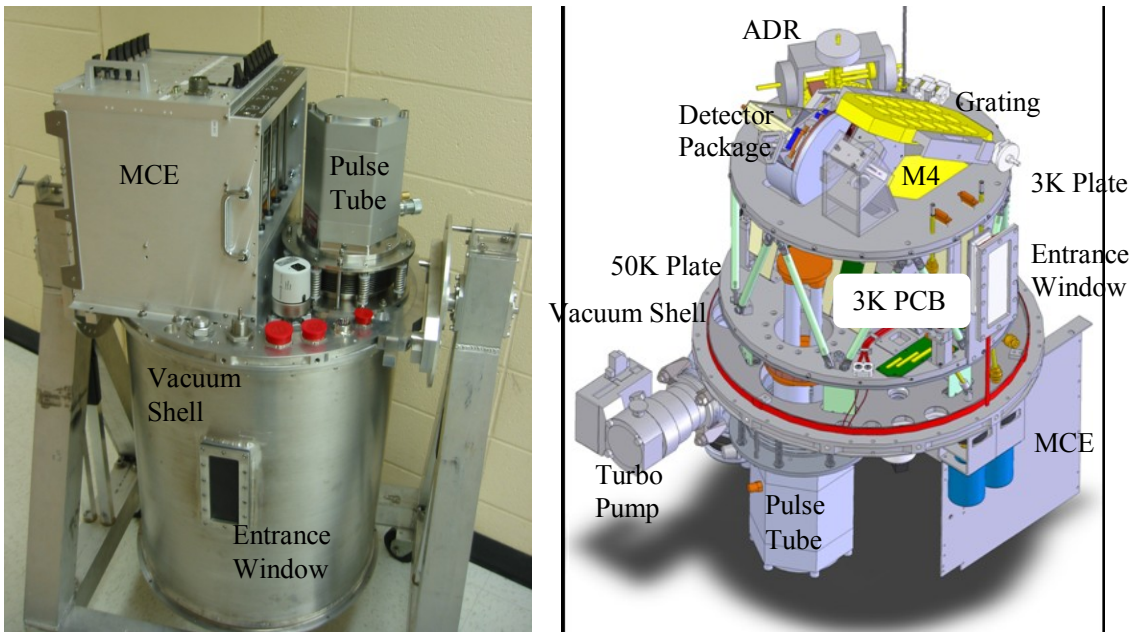

Figure 7: (left) Photograph of ZEUS-2 in the lab at Cornell. (right) A schematic of the ZEUS-2 internals.

Because of SQUID sensitivity to magnetic fields, adequate magnetic shielding is a priority. ZEUS-2 features a high-permeability shield (Amumetal – Amuneal Mfg. Co.) just inside the vacuum shell to shield against fields sourced externally. Inside the dewar the ADR is designed with several shields that significantly attenuate the field when the ADR magnet when it is in operation. However, additional shielding is needed for the SQUIDs to remain functional. All SQUIDs, including the entire detector package, are housed in a Cyroperm-10 shell surrounded by a superconducting niobium shell. This combination creates a zero field cavity around the SQUIDs once the niobium reaches it transition temperature of 7K and as long as the magnetic field remains below 2000 Gauss.

## 3.3 Detector Systems

### 3.3.1 Focal Plane Array

ZEUS-2 features a focal plane comprised of three arrays. All arrays feature transition-edge-sensed bolometers from Kent Irwin's groups at NIST Boulder. The arrays are manufactured on two silicon wafers with arrays 2 and 3 sharing one wafer. The usable FoV of ZEUS-2 is ~26 mm (spatial) x 57 mm (spectral) due to considerations of image quality. This usable FoV is split roughly equally between Array 1 and Array 2/3. The pixel sizes are 1.26, 1.06, and 2.06 mm square for Array 1, 2, and 3 respectively. For Array 1 and 2 there is a 0.14 mm structural/optical gap between each pixel (0.2mm for Array 3). Given that the final f/# = 2.75, this corresponds to 1.15 $\lambda/D$ pixels at 400 μm for Array 1. The array 2 pixels, however, are slightly oversized (1.8 $\lambda/D$) to compensate for point source flux that would be lost with diffraction limited pixels on APEX at 215 μm due to surface roughness (18 μm rms), and the challenges of pointing with a 4.4" diffraction limited beam at 215 μm. The nominal format (spatial x spectral) of Array 1, 2, and 3 are 9 × 40, 10 × 24, and 5 × 12 pixels respectively. The actual array designs have the corner pixels cut-out (as can be seen in Fig. 8) so that the total pixel count is 555. These cut-outs provide space for handling the arrays as well as improved structural stability. Little science is lost by these eliminating these pixels because poor image quality in these areas would have limited performance. Also, the spatial field is angled with respect to the spectral field and the cut-outs follow this angle.

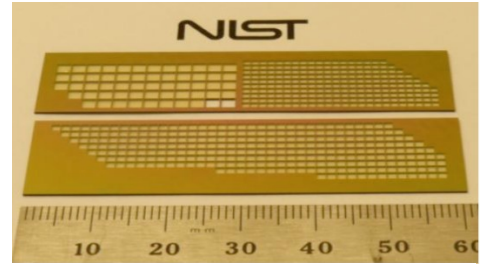

Figure 8: Structural mock-up of ZEUS-2 arrays. Array 1 is on the bottom. Array 2/3 is on top.

To achieve a high detective quantum efficiency (DQE) all of the arrays feature a tuned backshort and palladium-gold mesh absorber. A high DQE is desirable because it improves the background limited noise equivalent power (NEP), and makes it easier to achieve background limited performance. Array 1 is tuned to 400 μm ($\lambda/4$) resonance, and Array 2/3 is tuned to a 645 μm ($\lambda/4$) resonance. Modeling predicts that all arrays will operate with high quantum efficiency. Array 1 will operate with better than 90% DQE in both the 350 and 450 μm bands. Array 2 will see DQE > 90% in the 215 μm window ($3\lambda/4$ resonance), and Array 3 will see equally good DQE at the 645 μm windows ($\lambda/4$), but at 850 μm a DQE of ~78% is expected. The use of the mesh absorber is important as it allows for a high absorptive efficiency while adding very little to the heat capacity of the pixel. This is important if one wants to reach a low NEP while still maintaining a reasonable detector time-constant (< 10 msec). Modeling of the absorber at NIST Boulder has provided straight forward mesh designs for the longer wavelength arrays (see Fig. 9); however, the 215 µm required a more complex design. The final design features a mesh with 2 µm lines on a 42 µm pitch made of 12 Ω/square PdAu film giving an effective sheet impedance of 264 Ω/square—close to the ideal impedance of free space, 377 Ω/square,. By adding a capacitively coupled square ring to each hole in the mesh, the sheet impedance better matches free space as well as removes parasitic reactance resulting in devices with less than 3.2% reflection losses.

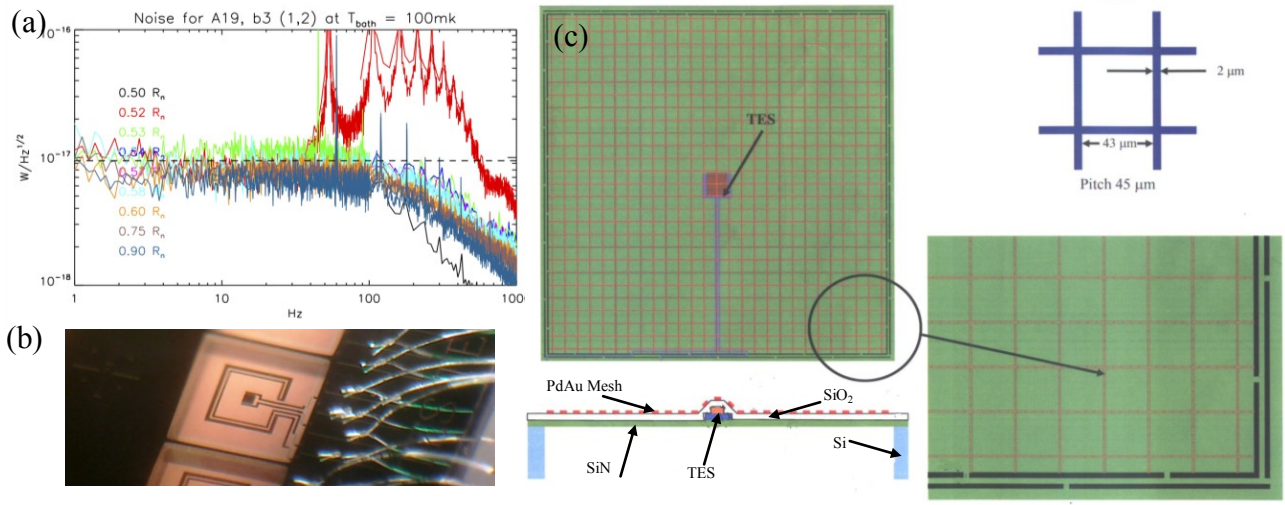

Figure 9: (a) Noise spectrum for a 400 micron array test pixel as of function of percent normal TES resistance. Noise levels match well with expected values. 53% and 52% $R_n$ noise levels are high due to electro-thermal oscillations that will be resolved in the next round of test pixels. (b) A dark-test pixel for the 215 micron array. The TES is the square in the middle. The loop around the TES is a heater used for measuring the thermal time constant of the bolometer. (c) Bolometer design for the 400 micron optical-test array. From the upper-left moving clockwise: 1.4 x 1.4 mm pixel with TES and mesh, PdAu mesh design, close-up of leg structure, and profile view of bolometer.

Recently, several dark test pixels (i.e. pixels without an absorber) were characterized. The goal of the characterization was to determine the pixels' thermal conductance, dark NEP, and thermal time constant, which when combined with G, gives the heat capacity of the bolometer. Figure 9 shows one of the silicon chips with test pixels and the noise spectrum of the pixel. Based on the results of these test a 400 micron optical test array is being fabricated for integration into ZEUS-2 in July 2010.

### 3.3.2 Detector electronics

ZEUS-2 utilizes the three stage time domain SQUID multiplexed readout system developed at NIST Boulder.[28] Each TES bolometer is coupled to one of the inputs on a 32 row SQUID multiplexors (20 MUX chips will be used in total). The multiplexors are housed in the detector package and their 1st stage SQUIDS (SQ1) are coupled to the 3rd stage SQUID series arrays (SQ3) by a 2nd stage SQUID (SQ2). The SQ3s are at 3K and amplify the detector signals by a factor of 100 before they reaching room temperature. All of the signal and bias are routed from room temperature to a 3K PCB over twisted-pair wires in a Tekdata cable. The 3K PCB also contains the SQ3s. The signal and bias lines are then sent to the detector package over copper-clad NbTi twisted pair Tekdata cable. The SQ2 to SQ3 interconnect uses two twisted pairs for each column to ensure that the inductance of that connection does not get too high and limit the multiplexing rate.

Because of the SQUID sinusoidal response to a linearly changing input signal, an active feedback scheme must be used to ensure the SQUIDs are operating in their linear regime. ZEUS-2 uses the Multi-Channel Electronics (MCE) developed for SCUBA-2 by Mark Halpern's group at the University of British Columbia to provide the active feedback.[29] The MCE also provides the SQUID bias to power each stage of SQUIDS, the passive feedback for the SQ2 and SQ3, row addressing, pixel heater biases, and the TES biases. The MCE reads the TES response by using a PI control loop to servo the SQ1 feedback, canceling any change in the output of the TES and keeping the SQ3 output constant. It is the SQ1 feedback that is then the TES response and indicates the power incidence on the bolometer.

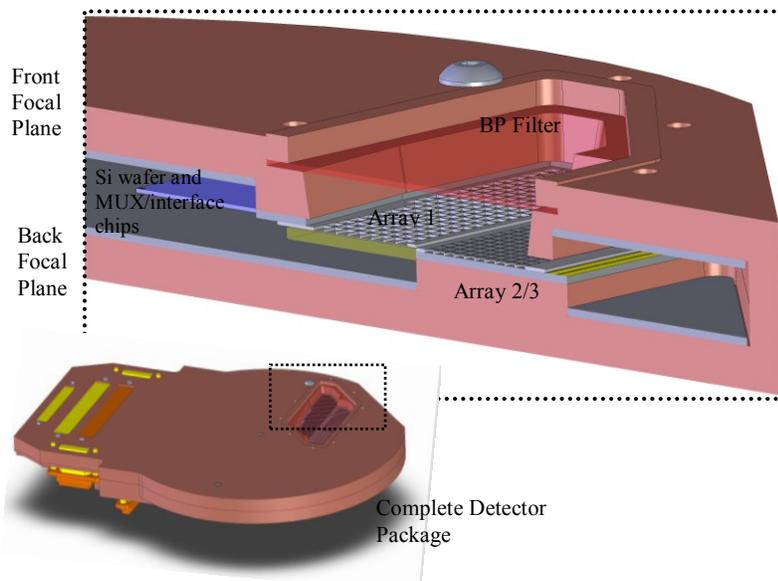

Figure 10: Detector package design. In this orientation the beam travels down from the top of the page. Array 1 is back illuminated and attached to the front plane while Array 2/3 is front illuminated and attached to the back plane. Each plane consists of a copper base, a Si wafer for MUX/interface chip mounting and wire routing, a raised platform for the array, a Si array carrier, and an array. The arrays sit on a raised platform so that the absorbers on the bolometers are coplanar. The band-pass filter mount is part of the front plane.

### 3.3.3 Detector Package

Developing a detector package that meets ZEUS-2's space limitations but at the same time fulfills the electronics requirements proved to be quite challenging. An initial design called for all 20 MUX chips, 20 interface chips (interface the TES with the MUC chips and provide the shunt resisters for TES voltage biasing), and all three arrays on one 6" silicon wafer. A silicon wafer chosen so that superconducting niobium traces could be patterned on the wafer to provide a superconducting interconnect between the arrays and the interface/MUX chips. Unfortunately there is not enough real-estate on a 6" wafer to fit all the necessary components and still leave room for traces and interfacing with cabling. A novel solution presented itself however; separate the arrays and mount Array 1 and Array 2/3 on separate wafers. This creates a front and back focal plan that are then sandwiched together as in show in Figure 10. This design has the added benefit of enabling ZEUS-2 to be easily operated with only one array installed. The arrays can also be easily swapped in and out for troubleshooting and upgrading of the bolometers.

## 4. SENSITIVITY

The noise equivalent power, NEP of a background limited spectrometer with warm optical elements at temperature T, emissivity ε, and cold optical elements of transmission τ (emissivity 0) and DQE η, is given by,[30]

$$NEP = h\upsilon \cdot \left\{ \frac{2A\Omega}{t_{int}\lambda^2} \cdot \Delta\upsilon \varepsilon \eta \tau \tilde{n} \cdot (1 + \varepsilon \eta \tau \tilde{n}) \right\}^{1/2} Watts\ Hz^{-1/2}. \quad (1)$$

Where $t_{int}$ is the integration time, h is Planck's constant, υ is the frequency, $\Delta\upsilon = \upsilon/R$ is the detection bandwidth, $A\Omega/\lambda^2$ is the number of photon modes, $\tilde{n} = 1(exp(h\upsilon/kT)-1)$ is the mode occupation number. The factor of 2 accounts for detecting both polarizations of light. The noise equivalent flux, NEF, is then NEF = 2·NEP/(A·η·τ·$t_w$), where the factor of 2 is due to chopping loss, and $t_w$ is the warm transmission of the system. If at the front of the dewar the warm transmission is just $t_w = \eta_{window}$, the transmission of the polyethylene window (92%). When considering an astronomical source, $t_w$ also includes the sky and telescope efficiencies, $\eta_{sky}(\lambda)$ and $\eta_{tel}(\lambda)$, as well as for point sources, the coupling to a pixel, $\eta_{pix}(\lambda)$.

Using equation 1 above, the expected point source sensitive for ZEUS-2 on CSO and APEX are listed for the band centers in Table 2 and plotted at all wavelengths in Figure 12. The current ZEUS sensitivities and the preflight sensitivity estimate for the Herschel instruments are also plotted. Because ZEUS-2 will be fully background limited and will have 1.28× better DQE, it is expected to be ~1.35 and 1.6 times more sensitive than ZEUS at 350 and 450 μm bands.

Table 2: ZEUS-2/CSO and ZEUS-2/APEX Band Center System Parameters and Sensitivity

| Band Center (μm) | Slit (") | R (λ/Δλ) | $\eta_{sky}$ | $\eta_{tel}$ | Power at Detector (Watts) | NEP at Detector (W Hz$^{-1/2}$) | $\eta_{pix}$ | NEF[a] (W m$^{-2}$ Hz$^{-1.2}$) | MDLF[b] (5σ, 4hrs, W m$^{-2}$) |
|---|---|---|---|---|---|---|---|---|---|
| **APEX:** | 0.2 mm precipital water vapor (PWV) for 200 – 300 μm bands, 0.5 mm at longer wavelengths | | | | | | | | |
| 202 | 5.6 | 1550 | 0.33 | 0.19 | 3.8E-12 | 1.2E-16 | 0.90 | 1.3E-16 | 3.8E-18 |
| 233 | 6.2 | 1600 | 0.33 | 0.25 | 2.3E-12 | 8.3E-17 | 0.86 | 7.5E-17 | 2.2E-18 |
| 295 | 6.9 | 1160 | 0.36 | 0.34 | 2.0E-12 | 7.3E-17 | 0.78 | 5.4E-17 | 1.6E-18 |
| 354 | 7.5 | 980 | 0.53 | 0.40 | 1.2E-12 | 5.3E-17 | 0.71 | 2.3E-17 | 6.8E-19 |
| 451 | 9.4 | 900 | 0.58 | 0.46 | 5.9E-13 | 3.3E-17 | 0.57 | 1.5E-17 | 4.4E-19 |
| 626 | 15 | 580 | 0.60 | 0.50 | 8.3E-13 | 3.5E-17 | 0.76 | 9.4E-18 | 2.8E-19 |
| 830 | 18 | 460 | 0.87 | 0.55 | 1.8E-13 | 1.2E-17 | 0.59 | 2.5E-18 | 7.5E-20 |
| **CSO:** | Assumes 0.72 mm PWV | | | | | | | | |
| 354 | 8.7 | 980 | 0.38 | 0.41 | 1.5E-12 | 5.9E-17 | 0.71 | 4.1E-17 | 1.2E-18 |
| 451 | 11 | 900 | 0.43 | 0.52 | 7.1E-13 | 3.6E-17 | 0.57 | 2.3E-17 | 6.9E-19 |
| 626 | 17 | 580 | 0.46 | 0.64 | 1.1E-13 | 3.8E-17 | 0.76 | 1.3E-17 | 3.9E-19 |
| 830 | 21 | 460 | 0.83 | 0.70 | 2.1E-13 | 1.3E-17 | 0.59 | 3.0E-18 | 8.9E-10 |

[a]The NEF refers to the point source detection, i.e. it includes $\eta_{pix}$, the coupling to a pixel. [b]MDLF is minimum detectable line flux (5σ) from a point source in 4 hours integration time.

The predicted ZEUS-2 sensitivities significantly exceed the sensitivities of all current systems, including Herschel. Except at 200 μm, ZEUS-2 is about a factor of 10 more sensitive than the Herschel spectrometers. At 200 μm ZEUS-2 is about equally sensitive to point sources as PACS, but has ~2.5 better spatial resolution. In figure 11 a ZEUS spectra is over plotted on a Herschel SPIRE spectra providing direct comparison of the two systems sensitivities. Currently, ZEUS has a equivalent heterodyne receiver temperature of $T_{rec} \sim 20$ K (single side-band), which is 4.6 and 2.3 times more sensitive than the best reported heterodyne values of 205 K (double side-band) and 250 K (single side-band) at 350 and 450 μm respectively when the effects of the background are included (a 60% emissive, 270 K blackbody)(K4 and G2).[31,32] Given its better sensitivity, we expect ZEUS-2 to be better than 3 and 5 times more sensitive than the best ground based heterodyne systems.

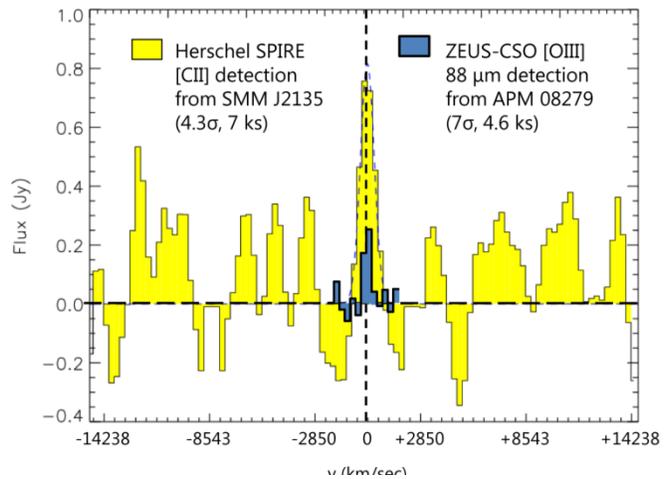

Figure 11: ZEUS-CSO spectrum[3] over plotted on a Herschel-SPIRE spectrum[19]. With velocity referenced to z = 2.327 and z = 3.911 for the Herschel and ZEUS spectra respectively.

## 5. SUMMARY

The goal of ZEUS-2 is to study the star formation history of the universe from just a few billion years after the big bang to the present epoch. Building on the work of ZEUS, we will (1) investigate star formation from z ~ 0.25 to 5 by measuring the redshifted fine-structure lines of distant galaxies, (2) measure redshifts from optically obscured galaxies by detecting the bright 158 μm [CII] line, and (3) investigate the properties of nearby starburst and ultra-luminous galaxies through their submillimeter fine structure and CO rotational lines.

ZEUS-2 is an ideal instrument for this work as the moderate resolution (R~1000) of the long slit echelle grating spectrometer is optimal for detection of faint, broad lines from distant galaxies. Using three state-of-the-art TES bolometer arrays that are tuned for optimal efficiencies at 215, 400, and 625 microns, ZEUS-2 achieves sensitivities that are unparalleled from ground our space. When complete, ZEUS-2 will be able to obtain spectrum from up to nine spatial positions on the sky and have access to seven different bands between 200 and 850 microns allowing ZEUS-2 to excel at both mapping of nearby systems, and redshift searches of distant sources.

At present we are integrating the optical and cryogenic systems of ZEUS-2. Our first optical array at 400 μm will be completed and integrated into ZEUS-2 by August 2010. Following in lab tests, science-grade arrays will be produced and integrated into ZEUS-2 in the fall of 2010 in preparation of our first observing run at CSO in early 2011 to be followed soon after with a trip to APEX.

**Acknowledgements:** ZEUS development and observations were supported by NSF grants AST-0096881 and AST-0352855. ZEUS-2 development is supported by NSF grants AST-0705256 and AST-0722220.

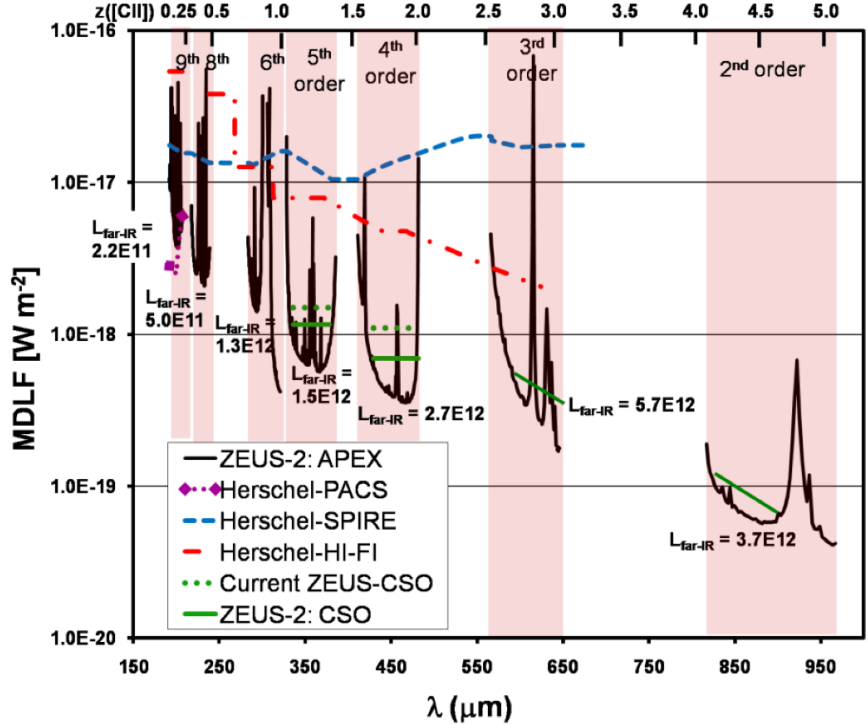

Figure 12: Minimum detectable line flux (W/m$^2$) (5σ, 4 hours) for ZEUS-2 on APEX and on CSO. We use telluric transmission models with 0.5 mm and 0.72 mm PWV for APEX and CSO (at band center) respectively— appropriate for good submillimeter nights at their sites (APEX windows < 300 μm use 0.2 mm). Also plotted is the sensitivity of ZEUS, the published preflight sensitivities of the 3 spectrometers on the Herschel telescope[33,34,35] with HIFI and PACS scaled to the resolving power of ZEUS-2: R = 1000, or Δυ = 300 km s$^{-1}$, appropriate for typical extragalactic line widths. ZEUS-2 is ~10 times more sensitive in the primary bands. The echelle orders, and the corresponding [CII] redshift intervals are listed, plus the minimum FIR luminosity galaxy detectable for each band assuming R= 0.1% as we measure for the z ~ 1 – 2 galaxies. We use $H_0$ = 71 km s$^{-1}$ Mpc$^{-1}$, $\Omega_0$ = 1, $\Omega_\Lambda$ = 0.7.